# Multiprocessor Scheduling of Dependent Tasks to Minimize Makespan and Reliability Cost Using NSGA-II

M.Rathna Devi and A.Anju

A.P/Department of IT, KCG College of Technology, Chennai, India

### ABSTRACT

*Algorithms developed for scheduling applications on heterogeneous multiprocessor system focus on a single objective such as execution time, cost or total data transmission time. However, if more than one objective (e.g. execution cost and time, which may be in conflict) are considered, then the problem becomes more challenging. This project is proposed to develop a multiobjective scheduling algorithm using Evolutionary techniques for scheduling a set of dependent tasks on available resources in a multiprocessor environment which will minimize the makespan and reliability cost. A Non-dominated sorting Genetic Algorithm-II procedure has been developed to get the pareto- optimal solutions. NSGA-II is a Elitist Evolutionary algorithm, and it takes the initial parental solution without any changes, in all iteration to eliminate the problem of loss of some pareto-optimal solutions.NSGA-II uses crowding distance concept to create a diversity of the solutions.*

### KEYWORDS

*Quality of Service, NSGA-II, elitism, makespan, reliability cost.*

## 1. INTRODUCTION

Generally, heterogeneous computing (HC) environments use a distributed collection of different high-performance machines, connected with high-speed links, for solving collections of computationally intensive problems that have different computational requirements [1, 2].Scheduling is the main problem that arises in heterogeneous computing environments. Scheduling means efficiently allocating a group of tasks coming into the system to the available resources in the system. However, finding optimal schedules in an HC environment is an NP-complete problem [3, 4]. In the HC environment considered, the tasks are assumed to be dependent, i.e., communications between the tasks are needed. The execution time of each task in each machine is already defined for the system. So static scheduling is done here. Static scheduling is utilized in different types of analysis and environments. The static scheduling is used for designing a system. Static scheduling is also used for post-mortem analyses i.e., to evaluate the performance of a dynamic scheduler, to check how effectively the system is using the resources available. Static scheduling is also used for simulation studies to know about the efficiency of hardware,while the system runs. In future, high-powered computational grids [5] will also be use static scheduling mapping techniques to distribute resources and computational power. Hence the wide applicability of static scheduling makes it an important area for ongoing research.

In the proposed work, each machine will execute only a single task at a particular time and no task will interrupt other task while executing,(i.e) no dependency among the tasks. The proposed work is done for non-preemptive tasks.The scheduling algorithms can be rated based on different parameters like makespan, flowtime, communication cost, reliability cost and makespan. This paper concentrates to minimize makespan and reliability cost. Makespan is the time taken for a HC system to finish the last task. Reliability is defined as the system should not fail during the time of executing the tasks. This scheduling problem proposed here is a multi-objective problem with the objective to minimize the makespan and reliability cost of the system.

In Genetic Algorithm, the parental characteristics are changed after the first generation by mutation. So, the time taken for obtaining sub-optimal solution is more. To solve this problem, elitism concept is used. Using elitism concept a new population is constructed to allow some of the better solutions from the current generation to carry over to the next, unaltered. Elitism is used to eliminate the problem of loss of good solutions by keeping the elite population. Crowding distance technique of NSGA-II is used to get diversity of the solution. In this paper, Non-dominated Genetic Algorithm (NSGA-II) is used to select the best optimal schedule by considering both the objectives and to get sub-optimal solution in minimum time.

The remainder of this paper is organized as follows. Section 2 provides existing methodologies. Section 3 defines the computational environment parameters. Description of NSGA-II based procedure is presented in Section 4. Section 5 presents the results for different problems simulated and finally, Section 6 concludes with finishing remarks and future work.

## 2. EXISTING WORK

Job scheduling is a tedious work in multiprocessor system than in a single processor system. Scheduling problem in multiprocessor system is always NP-hard. [6]. Now-a-days, more number of genetic algorithm (GA) are proposed. Mitra and Ramanathan proposed a GA for scheduling of non-pre-emptive tasks with precedence [7]. Lin and Yang developed a hybrid GA, where various operators are applied at a different stage of the lifetime, for scheduling non-pre-emptive tasks in a multiprocessor environment [8]. Monnier presented a GA implementation to solve a scheduling problem for real-time non-pre-emptive tasks [9]. However, these algorithms have only one objective such as minimizing cost, completion time or total tardiness. Oh and Wu presented a multi-objective GA for scheduling non-preemptive tasks in a soft real-time system with multiprocessors [10]. However, this algorithm did not refer to conflict between objectives, the so called Pareto optimum, and assume that the performance of all processors is the same. Theys et al. proposed a GA for static scheduling in a heterogeneous system [11]. Page and Naughton presented a GA for dynamic scheduling algorithm on a heterogeneous system [12]. Dhodhi presented a novel encoding method of GA for task scheduling on a heterogeneous system [13]. But, all these algorithms are designed for normal tasks without time constraints. Sandeep Jain and Shweta Makkar presented the scheduling of dependent tasks using Genetic Algorithm[14].kamaljit and Amit presented a heuristic based Genetic Algorithm to minimize execution time and throughput of dependent tasks[15]. Amanpreet Kaur1 and Prabhjot Kaur implemented mapping heuristic Genetic Algorithm to minimize Makespan and Schedule length

ratio[16].But these algorithms are designed without taking the initial population into consideration in all generations.

## 3. SYSTEM MODEL

The system is represented by Directed Acyclic Graph G= T, E where T represents a set of N tasks to be executed in the system and E is the directed edges represents the data communication between two tasks. Let $t_{i,j}$ be the execution time of task ti on machine mj, $1 \leq j \leq p$. It is assumed that the expected execution time $t_{i,j}$,E $1 \leq I \leq n$ and $1 \leq j \leq p$, is known. Let $e_{k,l}$ ε E indicate communication from task $t_k$ to $t_l$, where task $t_k$ ($t_l$) is said to be an immediate predecessor(successor)task of task $t_l$ ($t_k$).Associated with directed edge $e_{k,l}$ε E is the volume of data in terms of bytes ,which is denoted by $d_{k,l}$, that will be transmitted from task $t_k$ to task $t_l$ upon completion of task $t_k$.

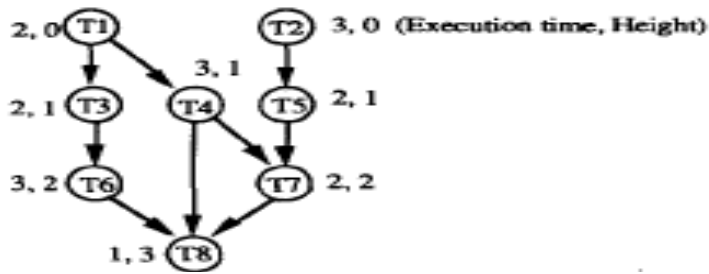

Fig 1. Directed Acyclic Graph

The scheduling problem for soft real-time tasks is formulated under the following assumptions: computation time of each task is known. The problem is multi-objective job scheduling considering, static scheduling of dependent jobs in heterogeneous environment, with the objective of completion time and reliability cost. Static scheduling means the tasks in the system are assigned to a particular machine before starting the system. So that there is no movement of task from one machine to another during the compilation time.

## 4. PROPOSED SOLUTION

Traditional optimization techniques and search do not take care more on problem domains and are not robust for multi-objective optimization problems. Previously Genetic Algorithm is used for static scheduling [17] in heterogeneous multiprocessor system. But GA could not produce a better pareto-optimal solution if more than one objective is considered. To overcome the above problem, it is proposed to use NSGA-II [18] in solving the problem.

### 4.1 NSGA-II

Initialize the population $P_t$.
Create the child population $Q_t$ from the previous population $P_t$.
Combine the two populations $Q_t$ and $P_t$ to form $R_t$.

$R_t = P_t \cup Q_t$

Find the all non-dominated fronts $F_i$ of $R_t$.
Initiate the new population $P_t+1 = 0$ and the counter=1
While $P_t+1 + F_i \leq$ Npop, do: $P_t+1 \leftarrow P_t+1 \cup F_i$, $i \leftarrow i+$
Arrange the last front $F_i$ in descending order using crowding distance and choose the first ($N_{pop} - P_t+1$) elements of $F_i$.
Use Genetic operators such as selection, crossover and mutation to obtain the new child population $Q_t+1$ size $N_{obj}$.

### 4.1.1 Initialization

Initialize the population $P_t$ using equality and inequality constraints. After initialization step, it creates chid population $Q_t$ from the currently existing population $P_t$ and then combines both populations to form $R_t$. Where $R_t$ is define as:[19]

$$R_t = P_t \cup Q_t$$

Before schedule creation the height of all tasks are found out. Then using permutations different possibilities of tasks in each height is found out. Then the tasks are allocated to the processor in order of this permutation combination results. A schedule would be illegal if a task is scheduled to be executed before its ancestor. Suppose that T and T' are tasks assigned to the same processor, and that T' is an ancestor of T. By the definition of height, we have height (T') < height (T). If we order the tasks in ascending order of height, then T' will be executed before T, and the schedule will be legal.

### 4.1.1.1 Algorithm

This algorithm randomly generates a schedule of the task graph TG for a multiprocessor system with p processors.

GSl. [Initialize.] Find height ht' for all task in task graph.
GS2. [Find the tasks based on their height and keep separately] Divide the tasks in task graph into different groups, G(h) (G(h) represents a group of tasks with height h), based on the value of ht'.
GS3.[Task permutation] Task within the particular height is get permutated to get different possibilities.
GS4. [Allocation to processors] From the different possibilities, allocate the jobs to processors p1 to pn. Each time start the allocation from subsequent processors.
By repeatedly applying the algorithm for all the possibilities, the initial population of search nodes can be generated.

### 4.1.2 Non-Dominated Sorting

After the initialization, the population is sorted based on non-domination. Each solution assigned a fitness value according to its non-dominated level, where level one is considered to be the best level. The solution at the level one did not dominate by any of other solution. Whereas solutions at other level dominated by least one solution. Perform the non dominated sorting to the initial population and identify the different rank to each population: rank1, rank2, rank3....etc. Rank provides the value of non-dominated rank across each solution. All the solutions have sorted according to its non-dominated rank.

```
for each (pεP)
for each (qεP)
if (p<q) then
S_p = S_pU{q}
else if (q <p) then
n_p = n_p+1
end
end
if (n_p= = 0) then
F_l = F_lU{p}
end
end
While ( F_i#Φ)
Q = Φ
for each (pεF_i)
for each(qεS_p)
n_q = n_q-1
If( n_q = 0) then
Q = Q U {q}
end
end
end
i = i+1
F_i = Q
End
```

### 4.1.2.1 Crowding Distance

To have difference in chromosomes, we have to find out crowding distance. Following algorithm is The crowding distance is calculated by the below algorithm.

```
l = |I|
set I[i]_distance = 0
end
for each objective m
I = sort(I,m)
I[1]_distance = I[l]_distance = α
end
for (I = 2 to (l-1))
I[i]_distance=I[i}_distance+I(k+1)m-I(k-1)m/f^max_m-f^min_m
end
I(k).m
```

### 4.1.3 Selection

After the chromosomes are sorted based on non-domination and with crowding distance found out, the selection is done using a crowded-comparison-operator (>n) and best solution is selected. All solution is having two criterias.
1. A Non-domination rank ($r_i$) in population
2. A local Crowding distance ($I[i]_{distance}$)
$i >_n j$
if ($r_i < r_j$) or
if ($r_i = r_j$)
and
($I[i]_{distance} > I[j]_{distance}$)

### 4.1.4 Crossover

The crossover operators play the most important part in evolutionary algorithm. This is done by selecting chromosomes from the parental generation and interchanging their genes, new chromosomes are obtained. By this, we can obtain better quality offspring that will enter the next generation and enable the search to be done on new regions of solution space not searched yet. There are many types of crossover operators in the evolutionary computation, which depends on the chromosome representation.

#### 4.1.4.1 Algorithm

This algorithm performs the crossover operation on two strings (A and B) and generates two new strings.[19]

C1. [Select crossover sites.] Randomly generate a number, c, between 0 and the maximum height of the task graph.
C2. [Loop for every processor.] For each processor Pi in string A and siring B, do-step C3.
C3. [Find the crossover points.] Find the final job $T_{ji}$ in machine $P_i$ that has height c, and $T_{ki}$ is the task following $T_{ji}$. That is, c =height' (Tji) < height' ( Tki ) and height' (Tji) are the same for all i.
C4. [Loop for every processor.] For each processor Pi in string A and string B, do step CS.
C5. [Crossover.] Using the crossover sites selected in step C3, exchange the bottom halves of strings A and B for each processor Pi.
Although the crossover operation is powerful, it is random in nature and may eliminate the optimal solution. Typically, its application is controlled by a crossover probability whose value is determined experimentally. Furthermore, we can always preserve the best solution found by including it in the next generation.

### 4.1.5 Mutation

Mutation is a GA operation that modifies one or more genes from their current value.This might result in completely new gene values being added to the gene pool. With this new gene values, it is able to arrive at better solution than the previous one. Mutation operator helps to prevent the result stopping at any local optima. Mutation is done based on mutation probability. This probability indicates how many chromosomes has to undergo mutation operation. This probability value should be very low. If it is set to high, the search will be a simple random search. The mutation operation is summarized in the following algorithm: [21]

#### 4.1.5.1 Algorithm

This algorithm performs the mutation operation on a string and generates a new string.[19]
MI. [Pick a task]. Randomly pick a task, Ti.
M2. [Match height.] Search the string for a task, Tj, with the same height.
M3. [Exchange tasks.] Create a new chromosome by interchanging the two jobs, Ti and Tj, in the schedule.

Typically, the frequency of applying the mutation operator is controlled by a mutation probability whose value is determined experimentally.

### 4.2 Completion Time

It is the time when the latest job will be finished. For a scheduling problem the value should be minimized [20].

$$\min\ f_1 = \max\{t_i^F\},$$
$$\text{s.t}\ t_i^E \leq t_i^S\ \forall i,$$
$$t_i^E \geq t_j^E + \sum_{m=1}^{M} c_{jm} x_{jm}, \tau j \varepsilon pre\ (\tau_i) \forall i,$$
$$\sum_{m=1}^{M} x_{im} = 1 \forall i,$$
$$x_{im} \varepsilon \{0,1\} \forall i, m.$$

In above equations, notations are de.ned as follows:
Indices:

i, j task index, i, j = 1, 2, . . . , N
m processor index, m = 1, 2, . . . , M.
Parameters:
G = (T ,E) : task graph
T = {$\tau_1$, $\tau_2$, . . . , $\tau_N$}: a set of N tasks
E = {$e_{ij}$ }, i, j = 1, 2, . . . , N, i ≠ j : directed edges
N : Number of jobs
M : Number of machines
$t_i$ : the ith task, i = 1, 2, . . . , N
$e_{ij}$ : precedence relationship between task $\tau_i$ and task $\tau_j$
$e_{ij}$=1,if there are precedence $i_p$ between task $\tau_i$ and $\tau_j$.
$e_{ij}$=0 otherwise
cim :computation time of task $\tau_i$ on mth processor
$d_i$ : deadline of task $\tau_i$
pre*($\tau_i$): set of all predecessors of task $\tau_i$
suc*($\tau_i$): set of all successors of task $\tau_i$

pre($\tau_i$) : set of immediate predecessors of task $\tau_i$
suc($\tau_i$) : set of immediate successors of task $\tau_i$
$t_iE$ : earliest start time of task $\tau_i$
$t_iF$ : finish time of task $\tau_i$
$t_iF=\min\{t_iS +c_im,',d_i\}$
where $c_im' = c_im|x_im$
$tS_i$ : real start time of task $\tau_i$

### 4.3 Reliability Cost

Reliability is defined as, the system will not fail when a task is under execution. Consider a heterogeneous system with M PEs, P={$P_1,P_2…P_M$}, and a DAG containing N nodes,{$u_1,u_2,...u_N$}. Let $t_i(x)$ be the computation time of node $u_i$ for PE $P_j$. Let $f_j$ be the failure rate of PE $P_j$. Let $g_{kb}$ be the failure rate of the communication link from $P_k$ to $P_b$. Let $w_{ij}$ be the volume of data that task $u_i$ needs to send to task $u_j$. $d_{kb}$ represents the delay to transmit a single length data from $P_k$ to $P_b$. Let $X_{ij}$ denotes a binary number whether task $t_i$ is assigned to $P_j$, for assigned, 0, for not assigned. Reliability Cost as follows[21]:

$$RC = \sum_{j=1}^{M}\sum_{i=1}^{N} f_j X_{ij} c_i t_j(i') + \sum_{k=1}^{M}\sum_{b=1}^{M}\sum_{i=1}^{N}\sum_{j=1}^{N} g_{kb} X_{ik} X_{jb} w_{ij} d_{kb}$$

$R_{ij}$, represents the reliability cost for task $t_i$ to be scheduled on Processor $P_j$. Let pred(x') be the set which includes all $t_i$'s predecessors, $R_{ij}$ can be expressed as follows:

$$R_{ij} = f_j c_i t_j(i') + \sum_{p \in Pred(i)}\sum_{k=1}^{M} g_{kj} X_{pk} w_{pi} d_{kj}$$

Therefore,

$$RC = \sum_{i=1}^{N}\sum_{j=1}^{M} X_{ij} R_{ij}$$

Thus, to maximize system reliability, the Reliability Cost is needed to be minimized. Reliability means how the system is working without fault when some group of task is assigned to it. Reliability cost says about how well the system when a group of tasks are assigned to it. When reliability is minimized the reliability cost will be maximized.

## 5. TEST RESULTS

Numerical tests are performed with a randomly generated task graph. P-Method [22] is used for the generation task graph. Directed Acyclic Graph is generated by random number concept using P-method. Element $a_{ij}$ of the matrix is equal to 1 if there is a predecessor relationship from $\tau_i$ to $\tau_j$ ; otherwise, $a_{ij}$ is equal to zero. An adjacency matrix is generated randomly with all its lower triangular and diagonal elements equal to zero. The elements above the diagonal elements of the adjacency matrix are verified by a Bernoulli process with parameter $\varepsilon$, which represents a success. When the Bernoulli trial is a success, then the element is assigned a value of 1, for a failure the element is given a value of 0. With this method, a probability parameter of $\varepsilon = 1$ creates a totally sequential task graph, and $\varepsilon = 0$ creates an inherently parallel one. Values of $\varepsilon$ that lie in between these two extremes generally produce task graphs that possess intermediate structures.

For the tasks' computation time and delay, random numbers based on exponential distribution and normal distribution is used as follows:

$cE_{im}$ = random value for exponential distribution with mean 5
$cN_{im}$ = random value for normal distribution with mean 5
$rE$ = random value for exponential distribution with mean $cE_i$
$rN$ = random value for normal distribution with mean $cN_i$
$dE_i = tE_i + \max\{cEi_m \ \forall \ m\} + rE +$ Communication time
$dN_i = tE_i + \max\{cN_{im} \ \forall \ m\} + rN +$ Communication time

where $cE_{im}$ and $cN_{im}$ is the execution time of task $t_i$ on processor $P_m$ based on exponential distribution and normal distribution, respectively. $dE_i$ and $dN_i$ is the deadline of task $t_i$ based on exponential distribution and normal distribution, respectively.

Data transmitted between the tasks are randomly generated between the values 1 and 10. Data matrix is represented as n×n matrix, where 'n' represents the number of tasks in the task graph. Data will be transmitted when there is a dependency between the tasks, otherwise in the matrix the data is represented as zero. In other words the data matrix is based on adjacency matrix.

For calculating the reliability cost, the processor failure rate and the communication link failure rate is necessary. The link failure rate is represented as a p×p matrix, where 'p' represents the number of machines. The processor failure rate is a single dimensional matrix, and the number of column is equal to the number of machines. Both processor failure rate and link failure rate is generated randomly between the values 0.0000075 and 0.0000125.

The parameters of GA were set to 0.9 for crossover (pC, ), 0.1 for mutation (pM, ), and population size (popSize) is taken as twice the number of jobs taken.

**CASE I - 2 Machines and 10 Jobs**

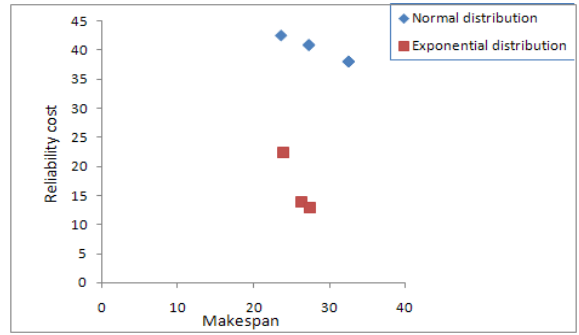

Fig. 2. Pareto optimal solutions for makespan for Normal and Exponential distribution(1 iteration)

The difference in the pareto-optimal solutions for computation time generated with the Normal and Exponential distribution for single iterations. With NSGA-II both the objectives get converged and both makespan and reliability cost get minimized.

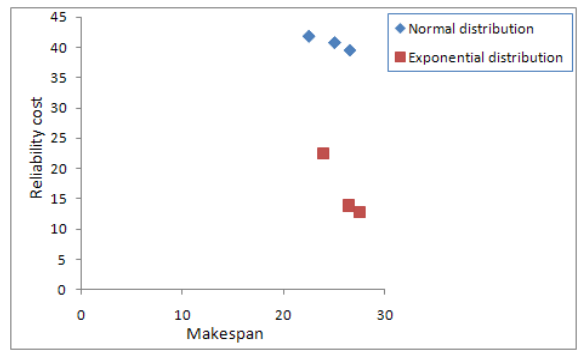

Fig. 3. Pareto optimal solutions for makespan for Normal and Exponential distribution(5 iteration)

The difference in the pareto-optimal solutions for computation time generated with the Normal and Exponential distribution for 5 iterations. When NSGA-II is used and number of iterations get increased both the objectives get converged and both the makespan and reliability cost get minimized.

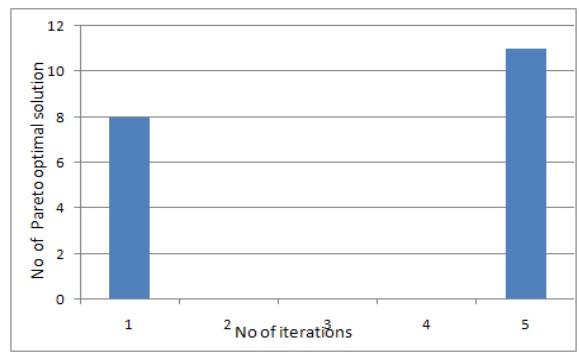

Fig 4. Pareto optimal solution for iteration 1 and 5(Normal distribution for makespan)

International Journal in Foundations of Computer Science & Technology (IJFCST), Vol.4, No.2, March 2014

The pareto -optimal solutions for Computation time generated with Normal distribution for different number of iterations and it is observed that more number of pareto-optimal solutions is obtained, when the number of iterations gets increased.

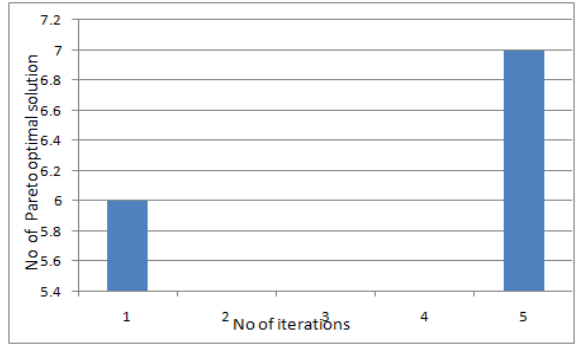

Fig 5. Pareto optimal solution for iteration 1 and 5(Exponential distribution for makespan)

The pareto optimal solutions for Computation time generated with Exponential distribution for different number of iterations and it is observed that more number of pareto-optimal solutions is obtained, when the number of iterations get increased.

**CASE II - 4 Machines and 50 Jobs**

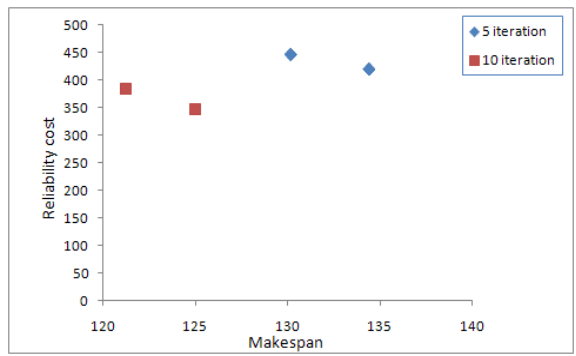

Fig 6. Pareto optimal solution for makespan for Normal distribution

The difference in the pareto optimal solutions for computation time generated with the Normal distribution for different iterations. When the number of iterations get increased both the objectives get converged and both the makespan and reliability cost get minimized.

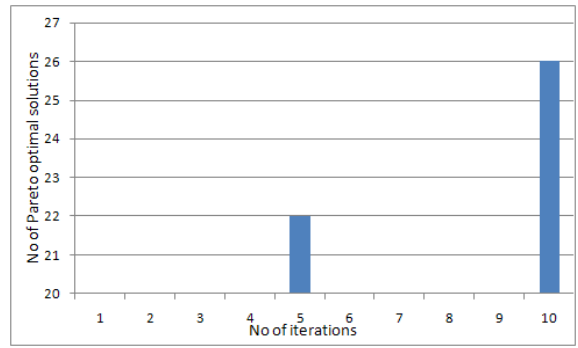

Fig 7. Pareto optimal solution for iteration 1 and 5(Normal distribution for makespan)

The pareto-optimal solutions for Computation time generated with Normal distribution for different number of iterations and it is observed that more number of pareto-optimal solutions is obtained, when the number of iterations get increased.

## 6. CONCLUSION

This paper gives solution for the scheduling problem in Multiprocessor system with multiple objectives to minimize the makespan and reliability cost. From the above work it is cleared that the deterministic methods are not efficient in solving multi-objective problems. The Non-dominated sorting Genetic algorithm-II based approach is a random search method that can be used to solve hard combinatorial optimization problems. The simulation was done for the computation time generated with Normal and exponential distribution. From the simulation results, it is found that the two objectives makespan and reliability cost is minimized successfully.

In future, study has to be done to select the correct parameter. It is planned to extend the proposed work for, dynamic scheduling of both dependent and independent tasks considering different objectives. It is planned to compare the simulated results with other scheduling algorithms also. It can also be extended by considering the objective like maximizing the processor utilization.

## REFERENCES

[1] S. Ali, T. D. Braun, H. J. Siegel, and A. A. Maciejewski, "Heterogeneous computing", Encyclopedia of Distributed Computing, Kluwer Academic, 2001.

[2] H.J. Braun et al, "A comparison of eleven static heuristics for mapping a class of independent tasks onto heterogeneous distributed computing systems" Journal of Parallel and Distributed Computing, 61(6), 2001.

[3] D. Fernandez-Baca, "Allocating modules to processors in a distributed system," IEEE Transactions on Software Engineering, Vol. SE-15, No. 11, Nov.1989, pp. 1427-1336.

[4] 0 . H. Sbarra and C. E. Kim, "Heuristic algorithms for scheduling independent tasks on nonidentical processors," Journal of the ACM, Vol. 24, No. 2, Apr. 1977, pp. 280-259.

[5] Foster and C. Kesselman, The Grid: Blueprint for a New Computing Infrastructure, Morgan Kaufman, New York, 1998.

[6] Yalaoui F, Chu C, "Parallel machine scheduling to minimize total tardiness", International Journal of Production Economics 2002;76(3):265–79.

[7] Mitra H, Ramanathan P.A, "Genetic approach for scheduling non-preemptive tasks with precedence and deadline constraints", in Proceedings of the 26th Hawaii international conference on system sciences, 1993. p. 556–64.

[8] Lin M,Yang L, "Hybrid genetic algorithms for scheduling partially ordered tasks in a multi-processor environment", in Proceedings of the sixth international conference on real-time computer systems and applications, 1999. p. 382–87.
[9] MonnierY, Beauvais JP, Deplanche AM, "A genetic algorithm for scheduling tasks in a real-time distributed system", in Proceedings of the 24th euromicro conference, 1998.
[10] Oh J,Wu C, "Genetic-algorithm-based real-time task scheduling with multiple goals", Journal of Systems and Software 2004;71(3):245–58.
[11] Theys MD, Braun TD, Siegal HJ, Maciejewski AA, Kwok YK, "Mapping tasks onto distributed heterogeneous computing systems using a genetic algorithm approach", in Zomaya AY, Ercal F, Olariu S, editors, "Solutions to parallel and distributed computing problems", New York:Wiley; 2001. p. 135–78 [chapter 6].
[12] Page AJ, Naughton TJ, "Dynamic task scheduling using genetic algorithm for heterogeneous distributed computing", in Proceedings of the 19th IEEE international parallel and distributed processing symposium, 2005. p. 189.1.
[13] Dhodhi MK,Ahmad I,Yatama A,Ahmad I, "An integrated technique for task matching and scheduling onto distributed heterogeneous computing systems", Journal of Parallel and Distributed Computing 2002;62:1338–61.
[14] Sandeep Jain,shweta Makkar, "Multiprocessor Environment using Genetic Algorithm", International journal of Advanced research in computer science and Software Engineering, 2012;p.131-134.
[15] Kamaljit Kaur, Amit Chhabra, "Heuristics Based Genetic Algorithm for Scheduling Static Tasks in Homogeneous Parallel System" vol. 4, pp.183-198.
[16] Amanpreet Kaur, Prabhjot Kaur, "Implementation of Mapping Heuristic Genetic Algorithm", International Journal of Advanced Research in Computer and Communication Engineering,2013;p.3224-3229.
[17] Javier Carretero, Fatos Xhafa and Ajith Abraham, "Genetic Algorithm Based Schedulers for Grid Computing Systems", International Journal of Innovative Computing, Information and Control 3, 6, 2007
[18] K. Deb, A. Pratab, S. Agarwal, and T. Meyarivan, "A fast and elitist multiobjective genetic algorithm: NSGA-II", IEEE Transactions Evolutionary Computation, vol. 6, pp.182–197, 2002.
[19] Edwin S . H . Hou ,"A Genetic Algorithm for Multiprocessor Scheduling"Member, IEEE, Ninvan Ansari, Member, IEEE.
[20] MyungryunYoo, "Scheduling algorithm for real-time tasks using multi-objective hybrid genetic algorithm in heterogeneous multiprocessors system", Mitsuo Gen Graduate School of Information, Production & Systems, Waseda University, 2-7 Hibikino, Wakamatsuku, Kitakyushu, 808-0135 Japan
[21] Yi He, Zili Shao, Bin Xiao, Qingfeng Zhuge, Edwin Sha, "Reliability Driven Task Scheduling for Heterogeneous Systems", Department of Computer Science University of Texas at Dallas Richardson, Texas 75083, USA
[22] Al-Sharaeh S, Wells BE, "A comparison of heuristics for list schedules using the box-method and P-method for random digraph generation",in Proceedings of the 28th Southeastern symposium on system theory, 1996. p. 467–71.

**Authors**

Rathna Devi received M.Tech degree in Information Technology from P.S.G College of Technology Coimbatore, India. Since July 2013, she has been working as an Assistant professor in KCG College of Technology, Chennai, India. Her research area is on Grid Computing and Evolutionary Algorithms.

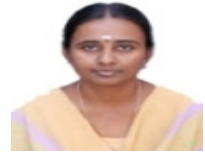

A.Anju received M.E degree in Computer Science Engineering from Sathyabama University, Chennai, India. Since July 2013,she has been working as an Assistant professor in KCG College of Technology, Chennai, India. Her research area is on Grid Computing and Evolutionary Algorithms.

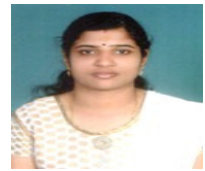